\documentclass[prd,twocolumn]{revtex4}

\pdfoutput=1

\usepackage{graphicx}
\usepackage{amssymb,amsmath,bm}  

\usepackage{lipsum}

\usepackage[usenames,dvipsnames]{xcolor}
\usepackage{hyperref}
\hypersetup{
    colorlinks = true,
    citecolor = {MidnightBlue},
    linkcolor = {BrickRed},
    urlcolor = {BrickRed}
}
\usepackage{lipsum}

\newcommand{\TIB}[1]{\textcolor{black}{#1}}
\newcommand{\TIBO}[1]{\textcolor{black}{#1}}

\newcommand{\uKam}{\mu\text{K-arcmin}}

\newcommand{\bl}{{\bm \ell}}
\newcommand{\bx}{{\bm x}}
\newcommand{\ba}{\begin{eqnarray}}
\newcommand{\ea}{\end{eqnarray}}

\begin{document}
\title{Measuring Polarized Emission in Clusters in the CMB S4 Era }
\author{ Thibaut Louis$^{1,2}$, Emory F. Bunn$^3$,  Benjamin Wandelt$^{1,4,5}$ , Joseph Silk$^{1,6,7,8}$ }
\affiliation{$^{1}$UPMC Univ Paris 06, UMR7095, Institut d'Astrophysique de Paris, F-75014, Paris, France\\
		$^{2}$ Laboratoire de l'Acc\'el\'erateur Lin\'eaire, Univ. Paris-Sud, CNRS/IN2P3, Universit\'e Paris-Saclay, Orsay, France\\
             $^{3}$ Physics Department, University of Richmond, Richmond, VA  23173, USA \\
             $^{4}$ Sorbonne Universit\'{e}, Institut Lagrange de Paris (ILP), 98 bis Boulevard Arago, 75014 Paris, France\\
             $^{5}$ Departments of Physics and Astronomy, University of Illinois at Urbana-Champaign, Urbana, IL 61801, USA \\
	    $^{6}$ AIM-Paris-Saclay, CEA/DSM/IRFU, CNRS, Univ Paris 7, F-91191, Gif-sur-Yvette, France \\
	    $^{7}$ Department of Physics and Astronomy, The Johns Hopkins University, Baltimore, MD 21218, USA \\
	    $^{8}$  BIPAC, University of Oxford,1 Keble Road, Oxford OX1 3RH, UK
             }
\begin{abstract}
The next generation of CMB experiments (CMB Stage-4) will produce a Sunyaev-Zel'dovich (SZ) cluster catalog containing $\sim10^{5}$ objects, two orders of magnitudes more than currently available. In this paper, we discuss the detectability of the polarized signal generated by scattering of the CMB quadrupole on the cluster electron gas using this catalog. We discuss the possibility of using this signal to measure the relationship between cluster optical depth and mass. We find that the area of observation of S4 maximizes the signal-to-noise (S/N) on the polarized signal but that this S/N is extremely small for an individual cluster, of order  0.5$\%$ for a typical cluster in our catalog, the main source of noise being the residual primordial E-mode signal.  However, we find that the signal could be detected using the full cluster catalog and that the significance of the result will increase linearly with the size of the CMB S4 telescope mirror. 
\end{abstract}

  \date{\today}
  \maketitle

\section{Introduction}\label{sec:intro}

Scattering of the cosmic microwave background (CMB) photons with the hot electron gas inside galaxy clusters generates a linear polarization
signal proportional to the CMB temperature quadrupole anisotropy at the cluster location. By measuring this polarized signal, we could in principle measure the \TIB{projected} quadrupole anisotropy as a function of position on the sky and redshift \cite{2014PhRvD..90f3518H,2006PhRvD..73l3517B,2000MNRAS.312..159C,2003PhRvD..67f3505C,2004PhRvD..70f3504P,2017arXiv170508907D, 2016MNRAS.460L.104L,2012PhRvD..85l3540A,1997PhRvD..56.4511K} and use it to constrain the $\Lambda$CDM model.
There are, however, many difficulties in such a measurement. First, the cosmological signal is intrinsically small, with a polarization amplitude of order 2 $\mu$K, one order of magnitude smaller than the polarized signal generated on the last scattering surface and two orders of magnitudes smaller than the temperature anisotropies. Second, this cosmological signal is modulated by the cluster optical depth, which is of order $\tau \sim10^{-2}$ at the center of a typical cluster, giving a typical polarized emission of  $2\times 10^{-2}\, \mu$K at the cluster location. Finally, the remote quadrupole at the cluster locations is highly correlated with our local CMB measurement, thus adding very little new information on the cosmological parameters in the standard $\Lambda$CDM model.

Without a dedicated, high-resolution, low-noise survey, the detection of this signal for individual clusters is out of reach in the near future. However, as pointed out by \cite{2014PhRvD..90f3518H,2017arXiv170508907D}, the next generation of CMB experiments could be sensitive to the polarized signal generated by a cluster population.

In this paper, we investigate the possibility of detecting this signal in the context of CMB Stage-4 (CMB S4 or S4) \cite{2016arXiv161002743A}, a next-generation CMB experiment composed of a set of ground-based facilities.
The aim of CMB S4 is to cover half the sky with noise levels in polarization of order 1.4 $\mu$K arcmin and at high angular resolution.  Such high sensitivity and large sky coverage is expected to increase the size of the corresponding cluster catalogue by two order of magnitudes compared to the currently available Planck cluster catalog \cite{2016A&A...594A..27P}.

We also discuss another way of using the signal. Instead of using it directly to measure cosmological parameters or test the $\Lambda$CDM model, we propose to calibrate the relationship between cluster mass and optical depth using the polarized emission. A simple relationship between optical depth and cluster mass has been pointed out recently using hydrodynamical simulations of clusters \cite{2016JCAP...08..058B},  but the exact parameters describing this relationship are unknown and depend on assumptions about the baryonic physics. A measurement of the cluster polarized emission could be used to calibrate the $\tau$-$M$ relationship.

In this approach, we exploit the high degree of correlation between the remote quadrupole measurement and our local CMB measurement. The very high signal-to-noise measurements of the largest modes of our last scattering surface by WMAP \cite{2013ApJS..208...19H} and Planck \cite{2016A&A...594A...1P} can be used to infer the expected cosmological signal at low and intermediate redshift, and a comparison between the expected polarized emission and the observed polarized emission allows us to constrain the cluster optical depth. 

Understanding the scaling of optical depth with cluster mass can be crucial for the interpretation of measurements of the kinetic Sunyaev-Zel'dovich (kSZ) effect. The kSZ effect produces secondary anisotropies on the temperature map  $T_{\rm kSZ}(\hat{n}) \propto \tau(\hat{n}) \bm{v}\cdot\hat{n}/c$ which are proportional to the cluster optical depth. The kSZ effect will be detected with very high significance with CMB S4 \cite{2016PhRvD..94d3522A}; however, the cosmological information encoded in the velocity field will be contaminated due to uncertainties on the optical depth.  The cluster polarization signal can be seen as an independent way to measure $\tau(\hat{n})$. The combination of the two measurements can be used to extract  cosmological information while reducing the contamination from  baryonic effects. \TIB{X-ray observation of clusters could also be used to measure cluster optical depth  \cite{2017ApJ...837..124F}. The method proposed in our paper is complementary to this approach and is affected by different observational and model systematics. It is also free from  selection effects.}

The paper is structured as follows. In Section \ref{sec:Pol}, we review the formalism and compute the expected polarized signal generated from the scattering of remote quadrupoles in the proposed CMB-S4 patch of observation. In Section \ref{sec:Model}, we describe our cluster model and compute the signal-to-noise per cluster of the CMB S4 cluster catalog. In Section \ref{sec:Forecast}, we show constraints on a power-law parametrization of the $\tau$-$M$ relationship for different experimental specifications of CMB S4. We discuss our results and conclude in Section \ref{sec:conclusion}. Technical details and beyond CMB S4 forecasts may be found in the appendices. 
We adopt the Planck fiducial cosmology \cite{2016A&A...594A..13P} with $\Omega_{\Lambda}= 0.685$,
$\Omega_{b}=0.049$, $\Omega_{m}=0.315$, $H_{0}=67\,
{\textrm{km s}^{-1}\textrm{Mpc}^{-1}}$, $n_s=0.96$ $A_s=2.2\times10^{-9}$,
and  optical depth at reionisation $\tau_{\rm reio}=0.06$.
Cluster masses $M_{500}$ are defined as the mass measured within
a radius $R_{500}$ that encloses a mean density $500$ times larger than the
critical density at the cluster redshift. 
\section{Polarization signal }\label{sec:Pol}
A detailed computation of the expected polarization signal due to remote quadrupole scattering is presented in \cite{2014PhRvD..90f3518H}. In this section, we summarize these results and discuss the expected signal in the CMB S4 patch of observation.
We denote ${\cal P}(\hat{n},z)= \tau(\hat{n},z) p(\hat{n},z)$ the polarized emission generated by the scattering of remote quadrupoles on clusters of optical depth $\tau(\hat{n},z)$ at redshift $z$, and focus on the calculation of the cosmological signal $p(\hat{n},z)$.
This signal can be decomposed into a part correlated with our measurement of the CMB temperature and an uncorrelated part:  $p(\hat{n},z)=p_{c}(\hat{n},z)+ p_{u}(\hat{n},z)$. It is the correlated part of the emission $p_{c}(\hat{n},z)$  that will be used to calibrate the $\tau$-$M$ relationship.

\subsection{Formalism }\label{subsec:form}

\begin{figure}
  \centering
  \includegraphics[width=1\columnwidth]{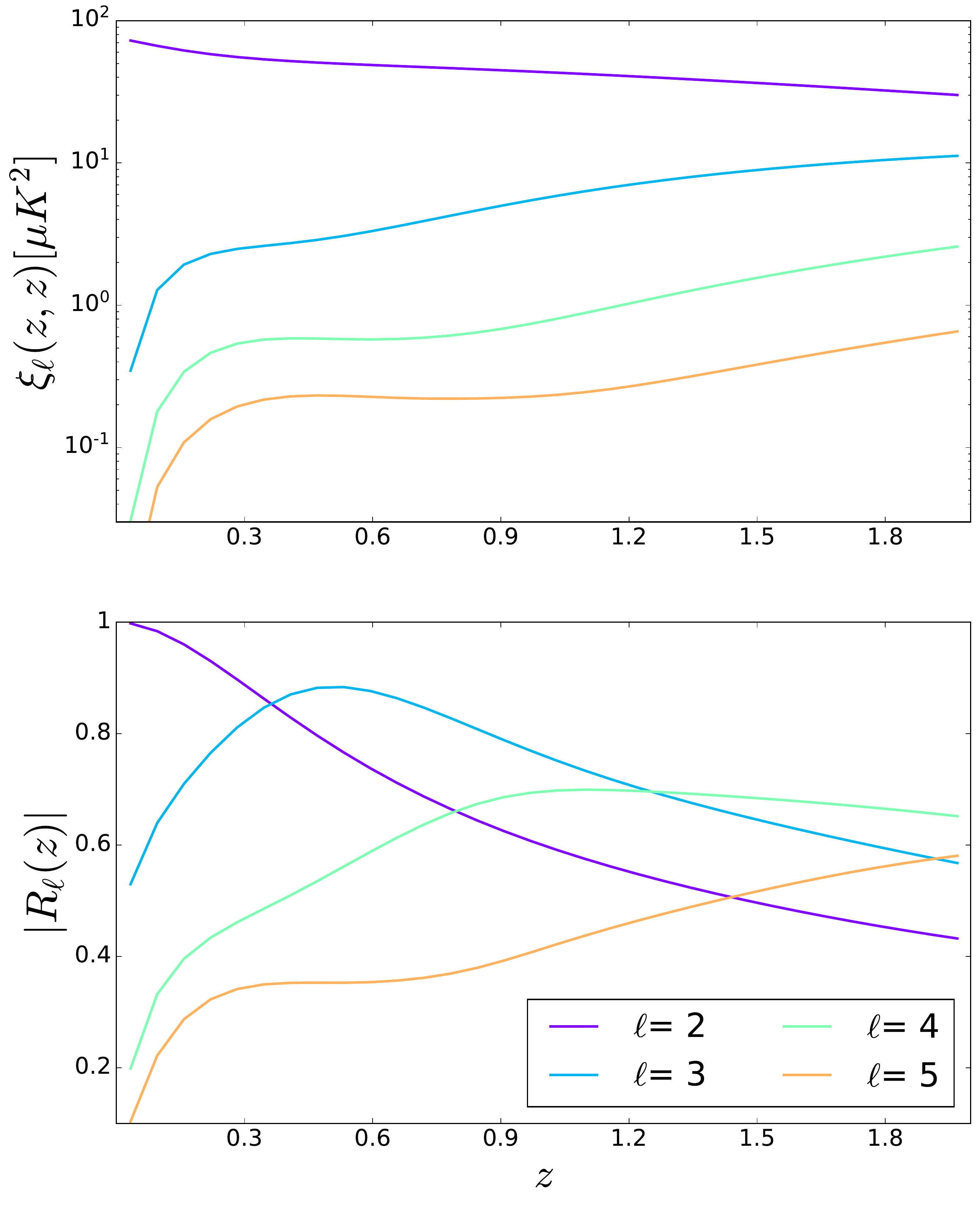}
  \caption{ \textit{Top}: power spectrum of the polarization field generated by remote quadrupole scattering. The field is purely quadrupolar at low redshift, but contributions of higher multipoles become relevant at high redshift.  \textit{Bottom}: correlation coefficient between this polarization field and our observations of the CMB. At low redshift, measurements of our own last scattering surface anisotropies allow us to infer the polarized emission with high accuracy.}
  \label{fig:corr}
\end{figure}

The complex polarization field $p(\hat{n},z)$ generated by remote quadrupole scattering can be expanded in spin-2 spherical harmonics with harmonic coefficients \cite{2014PhRvD..90f3518H}
\ba
&&(Q \pm iU)(\hat{n}, z) = \sum_{\ell m} p_{\ell m} (z)  [_{\mp 2} Y_{\ell m} (\hat{n})] \\
p_{\ell m} (z) &=& - i^{\ell}  3\pi \sqrt{f_{\ell}}  \int\frac{d{\bm{k}}}{(2\pi)^{3/2}} \frac{j_{\ell}(kr)}{(kr)^{2}} \Delta_{2}(k,r)\phi(\bm{k})Y^{*}_{\ell m}(\bm{k}) \nonumber
\ea
Here $Q$ and $U$ are the Stokes parameters describing the norm and orientation of the polarization field,  $f_{\ell}=\frac{(\ell+2)!}{(\ell-2)!}$ is a normalisation factor, and $\Delta_{2}(k,r)$ is the quadrupole transfer function, relating the gravitational potential $\phi(\bm{k})$ to the temperature quadrupole seen by an observer at a comoving distance $r(z)= \eta_{0}- \eta(z)$. The transfer function for small multipoles can be approximated as the sum of the Sachs-Wolfe and integrated Sachs-Wolfe effects,
\ba
\Delta_{\ell}(k,r) &=& \frac{1}{3}j_{\ell}[k(\eta-\eta_{*})] \\
&+& 2 \int_{\eta^{*}}^{\eta} d\eta' j_{\ell}[k(\eta-\eta')] \frac{\partial}{\partial \eta'} \left [\frac{D(\eta')}{a(\eta')} \right ], \nonumber 
\ea
where $\eta_{*}$ is the conformal time at decoupling, $D(\eta)$ is the growth factor, and we safely neglect the contribution coming from the Doppler effect. 

\begin{figure*}
  \centering
  \includegraphics[width=1\textwidth]{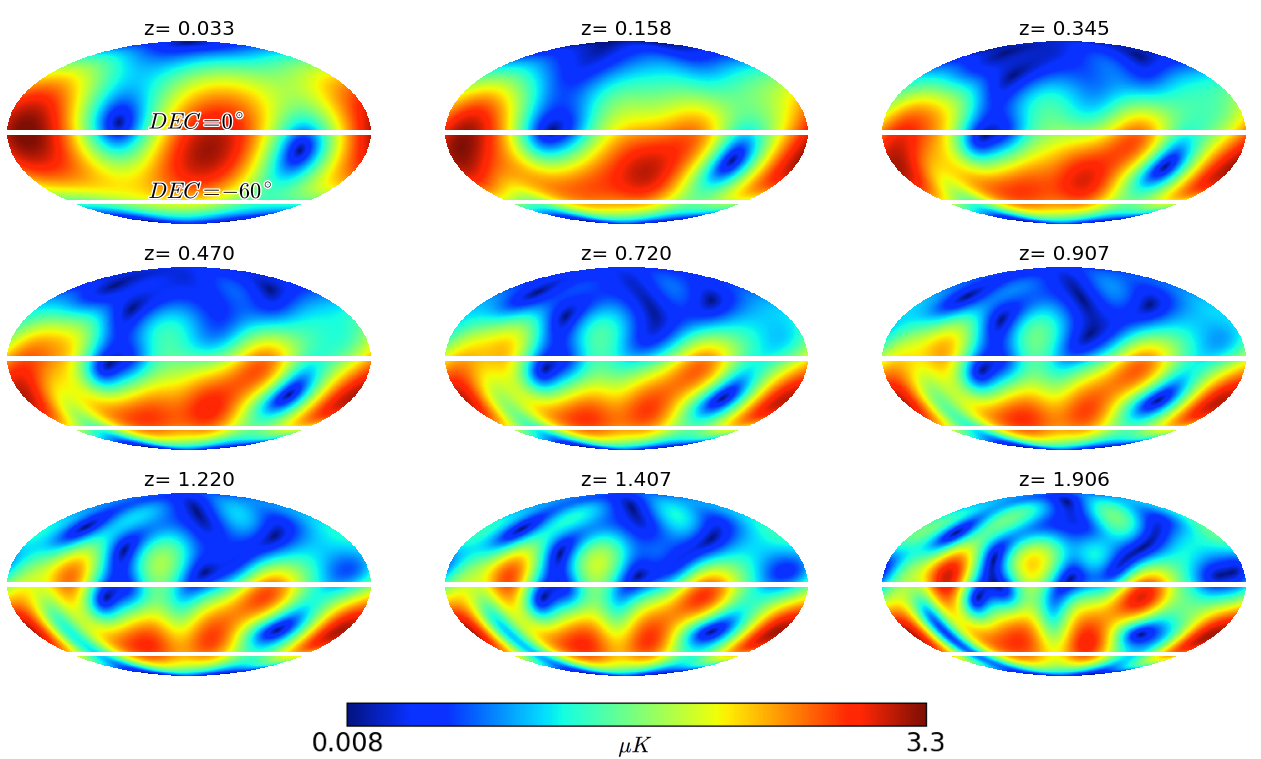}
  \caption{Absolute value of the polarized emission predicted from our local measurement of the CMB in equatorial coordinates. The white line at $\textrm{DEC}=0^{\circ}$ and $\textrm{DEC}=-60^{\circ}$ encompass the part of the CMB S4 area of observation that overlaps with LSST. At low redshift the signal is mostly quadrupolar, with smaller scales contributing at high redshift. The orientation of the large-scale modes of our last scattering surface results in most of the signal being located in the Southern hemisphere. The full polarized emission from remote quadrupole scattering will also include contribution from modes uncorrelated with our local measurement of the CMB. The importance of this uncorrelated signal will increase with redshift. }
  \label{fig:maps}
\end{figure*}

The power spectrum of the polarization field generated in two redshift slices $r(z)$ and $r(z')$ is given by 
\ba
 \xi_{\ell}(r,r')   &=&  \langle p_{\ell m} (r) p^{*}_{\ell m} (r') \rangle  \\
 &=& \frac{81 \pi f_{\ell}}{100} \int  \frac{dk}{k} \frac{j_{\ell}(kr)}{(kr)^{2}}  \frac{j_{\ell}(kr')}{(kr')^{2}}  \Delta_{2}(k,r) \Delta_{2}(k,r') P_{{\cal R}}(k) \nonumber
\ea
 $P_{{\cal R}}(k)$ is the dimensionless primordial curvature power spectrum.
 The top panel of Figure \ref{fig:corr} displays  the auto-power spectrum $\xi_{\ell}(z,z) $ as a function of redshift. At low redshift, the polarized emission is purely quadrupolar, but the contribution from higher multipoles increases as we  go to high redshift. 
\TIB{Note that this set of equations assumes that only super-horizon modes contribute, which is a good approximation for the signal of interest.}
 
The next step is to compute the correlation between this polarization field  and our observation of the CMB temperature anisotropies,  
\ba
 \zeta_{\ell}(r)   &=&  \langle p_{\ell m} (r) a^{*}_{\ell m}  \rangle \\
 &=& \frac{\TIB{-27}\pi \sqrt{f_{\ell}}}{25}  \int  \frac{dk}{k} \frac{j_{\ell}(kr)}{(kr)^{2}}  \Delta_{2}(k,r) \Delta_{\ell}(k,0)  P_{{\cal R}}(k). \nonumber  
\ea
The bottom panel of Figure \ref{fig:corr} shows the scaling of the correlation coefficient $R_{\ell}(z)= \zeta_{\ell}(z) / \sqrt{ \xi_{\ell}(z,z)C^{TT}_{\ell}}$ with redshift.  The interpretation of $R_{\ell}(z)$ is simple: it measures the accuracy with which we can predict the polarized emission using our local measurement of the CMB. 

\subsection{Polarized emission in the CMB S4 patch}\label{subsec:polS4}

Figure \ref{fig:corr} shows that measurements of the first few $a_{\ell m}$ of our observed CMB are sufficient to infer the part of the polarized signal correlated with our CMB observation, $p_{c}(\hat{n},z)$,  with high accuracy, and that the uncorrelated part of the emission is always sub-dominant at low redshift.

We compute the $a_{\ell m}$ using the spherical harmonic decomposition of the Planck SMICA temperature map \cite{2014A&A...571A..12P}. Using this map ensures that foreground contaminations which could affect the measurement of the largest angular scales of the CMB are minimal. \TIB{ The noise on the measurement of the first few $a_{\ell m}$  is  subdominant and is neglected in this analysis.}

The correlated part of the polarized emission is given by
\ba
p_{c} (\hat{n},z)=  \sum_{\ell m} \frac{\zeta_{\ell}(z)}{C^{TT}_{\ell}} a_{\ell m} {} [_{-2}Y_{\ell m}(\hat{n})]
\ea
We display in Figure \ref{fig:maps} the norm of the correlated part of the polarization field $\sqrt{ |p_c(\hat{n},z)|^{2}}=\sqrt{Q_{c}^{2}(\hat{n},z)+U_{c}^{2}(\hat{n},z)}$ in equatorial coordinates as a function of redshift.  As expected the signal is on the largest scales at low redshift and gets contributions from smaller scales at high redshift. 

 \begin{figure}
  \centering
  \includegraphics[width=1\columnwidth]{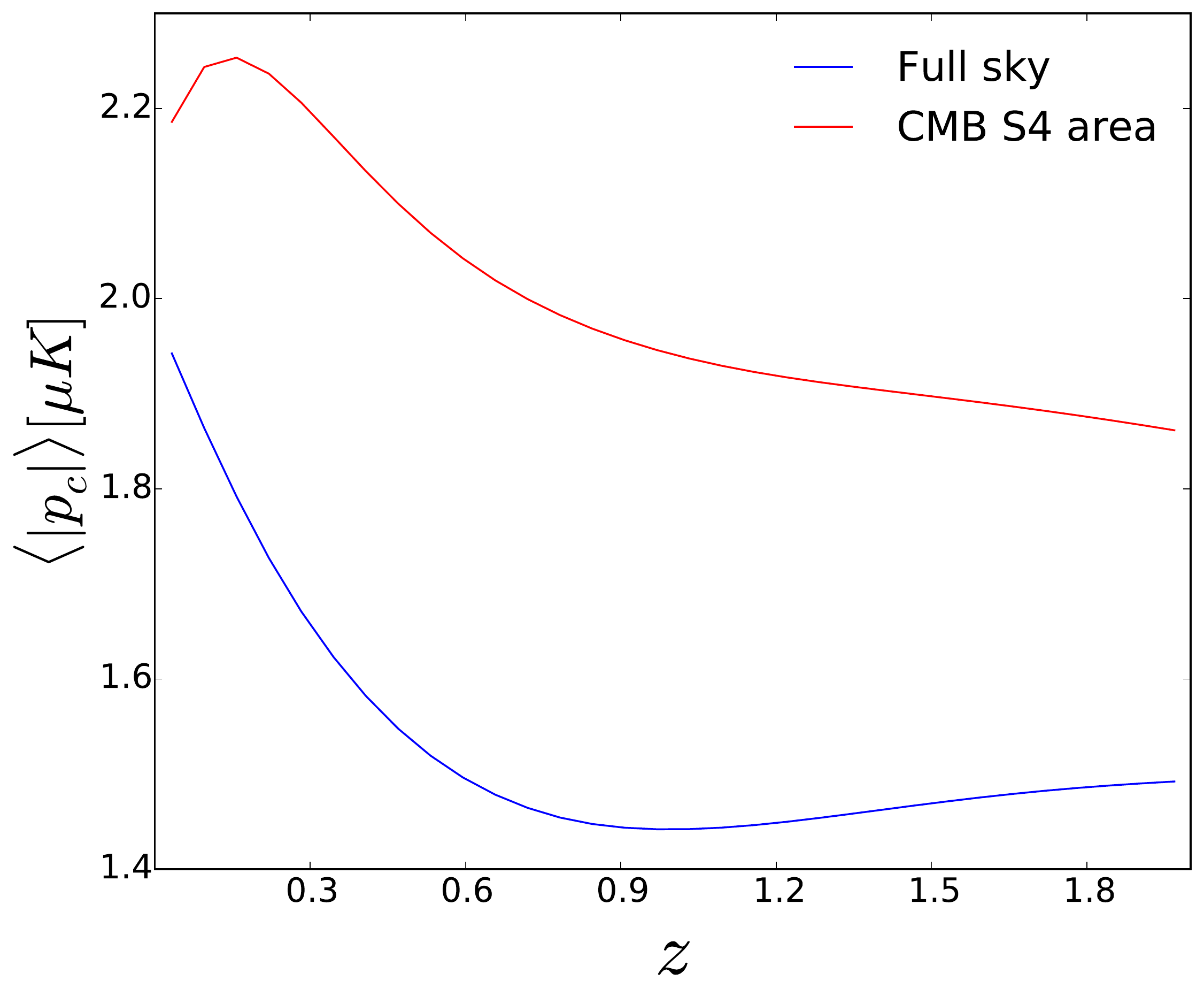}
  \caption{ \TIB{Sample} mean of the absolute value of the correlated part of the polarized emission.  The  emission in the S4 area of observation is significantly higher than the full-sky average. }
  \label{fig:average}
\end{figure}

CMB S4 \cite{2016arXiv161002743A} will be composed of a set of ground-based telescopes observing the sky from both Chile and the South Pole but does not yet include a component in the Northern Hemisphere. Consequently, we limit the S4 patch of observation to pixels lying below $\textrm{DEC}=0^{\circ}$. This is a conservative limit: low-elevation scans from Chile have been shown to reach $+20^{\circ} \textrm{DEC}$ \cite{2016SPIE.9910E..14D}. We do not exclude the part of the sky contaminated by the Milky-Way emission in this analysis, assuming that multi-frequency coverage will allow for partial foreground cleaning and that the remaining foreground emission will be sub-dominant with respect to other sources of errors.  We also use a lower limit of   $\textrm{DEC}> -60^{\circ}$, ensuring that the usable S4 patch of observation has overlap with the LSST telescope \cite{2012arXiv1211.0310L}. LSST could then be used to measure the redshift of clusters on which remote quadrupoles scatter. The slow evolution of the cosmological signal with redshift ensures that photometric redshift errors will not be an important contribution in the error budget of this analysis.

In summary, we will use for our forecasts the region delimited by the two white lines in Figure \ref{fig:maps}, accounting for approximately $40\%$ of the sky.

Because the signal is dominated by few modes on very large scales, the orientation on the sky of these modes matters. It is clear from Figure \ref{fig:maps} that most of the measurable signal at intermediate redshift is conveniently located in the Southern hemisphere, accessible by CMB S4. The difference between the average correlated polarized emission in the S4 patch of observation and the full-sky average is illustrated in Figure \ref{fig:average}.

\section{Cluster model}\label{sec:Model}
In this section, we introduce the physical model describing the emission of a single cluster.  We discuss the expected number of clusters detected through their thermal Sunyaev-Zel'dovich (tSZ) emission for the CMB S4 experiment and quantify their associated polarized emission. We forecast the signal-to-noise on the future measurement of the cluster polarization signal as a function of cluster mass and redshift.

\subsection{Cluster emission}\label{subsec:profile}

We first consider the tSZ effect which will be used to detect cluster and construct the CMB S4 cluster catalog.  The tSZ effect accounts for the inverse Compton scattering of CMB photons with the hot electron gas inside clusters producing secondary temperature anisotropies \cite{1980ARA&A..18..537S}
\ba
 \left . \frac{\Delta {\rm T}}{{\rm T}} \right |_{\rm tSZ}(\nu,\hat{n}) &=& f_{\rm tSZ}(\nu)\frac{\sigma_T}{m_ec^2}\int P_e(l,\hat{n})\, dl \\
 &=& f_{\rm tSZ}(\nu)\frac{\sigma_T}{m_ec^2}\int n_{e}(l,\hat{n})k_BT_e(l,\hat{n})  dl \nonumber
\ea
Here $\sigma_{T}$ is the Thomson scattering cross section, $P_e$, $n_{e}$ and $T_{e}$,  are respectively the electron pressure, the electron density, and the electron temperature, and  $f_{\rm tSZ}(\nu)$ accounts for the dependence of the effect with frequency.

We now consider the polarized emission generated by remote quadrupole scattering, which can be written
\ba
  {\cal P}  (\hat{n}) &=& \int p(l, \hat{n})\tau(l,\hat{n})\, dl  \\
 &=& \sigma_{T} \int p(l, \hat{n})n_{e}(l,\hat{n})\, dl \nonumber \\
 & \approx &  p \sigma_{T} \int n_{e}(l,\hat{n})\, dl \nonumber
\ea
Here $p(l, \hat{n})$ is the cosmological signal defined in Section  \ref{sec:Pol}.  We extract $p$  from the integral because the coherence length of the signal is much larger than the cluster size. 

We define the tSZ and optical depth amplitude  as
\ba
a_{\rm tSZ}  &\equiv&  \frac{4\pi \sigma_T}{d^{2}_{A}(z)}  \int_0^{R_{500}}dr r^2 n_e(r)\frac{k_BT_e(r)}{m_ec^2}= Y_{500} \\
a_{\tau} &\equiv&  \frac{4\pi \sigma_T }{d_A^2(z)}\int_0^{R_{500}}dr\,r^2\,n_e(r) = \tau_{500}
\ea
where $d_{A}(z)$ is the angular diameter distance at redshift $z$.   We use the GNFW/Arnaud profile \cite{2010A&A...517A..92A} to describe the tSZ pressure profile and the cored NFW model to describe the electron density profile \cite{2014PhRvD..90f3518H}
\ba
\Gamma_{\rm tSZ}(x) &=&\left[(xc_{500})^{\gamma}[1+(xc_{500})^{\alpha}]^{(\beta-\gamma)/\alpha} \right]^{-1} \\
\Gamma_{ \tau }(x) &=& \left[ (x+x_{0})c_{500}(1+c_{500}x)^{2}  \right]^{-1}
\ea
where $x$ is the dimensionless radial variable $x=r/R_{500}$ and the best-fit values of the GNFW/Arnaud profile $\alpha=1.062$, $\gamma=0.3292$, $\beta=5.4807$, $c_{500}=1.156$ are taken from \cite{2010A&A...517A..92A}. The core parameter $x_{0}=0.02 R_{200}/R_{500}$ is assumed to be fixed for all clusters of the catalog \cite{2014PhRvD..90f3518H}.
Using this notation, the emission centered a single cluster can be written
\TIB{
\ba
\delta T(\nu, \theta) &=& a_{\rm tSZ} f_{\rm tSZ}(\nu) g_{ \rm tSZ} (\theta/\theta_{500})\\
   {\cal P} (\theta) &=& p a_{\tau}   g_{\tau} (\theta/\theta_{500})
\ea}
with
\ba
g_{ \{\tau, \rm tSZ \}}(x) &=&   \frac{\int_{-\infty}^{\infty} dx_{z} \Gamma_{ \{ \tau, \rm tSZ \} } ( \sqrt{x_{z}^{2} + x^{2}}) }{ 4\pi \theta^{2}_{500} \int^{1}_{0}dx_{r} x^{2}_{r}\Gamma_{ \{ \tau, \rm tSZ \}}(x_{r}) }.
\ea
We do not model the kinematic Sunyaev-Zel'dovich (kSZ) effect arising due to the bulk motion of the cluster. It is subdominant compared to the tSZ effect and has a different frequency scaling. We also do not consider the polarized emission due to the transverse velocity of the cluster. Not including it contributes to negligible bias and increased variance for the recovery of the polarized signal generated by remote quadrupole scattering \cite{2014PhRvD..90f3518H}.

\subsection{Matched filter}\label{subsec:pol}

The maximum-likelihood solution for the amplitudes $a_{\rm tSZ}$ and $a_{\tau}$ can be obtained using matched filtering of the data. A matched filter uses  knowledge of the spatial profile to optimally recover the amplitude while suppresssing other components
with different spatial/spectral distributions \cite{2002MNRAS.336.1057H,2006A&A...459..341M,2004A&A...420...49F}.
We can write a data model for the cluster emission at frequency $\nu$ in a small patch around a cluster 
\TIB{
\ba
\delta T_{\nu}(\bx) &=&\delta  T_{\rm CMB} (\bx) + f_{\rm tSZ}(\nu) g_{\rm tSZ}(\bx) a_{\rm tSZ}+ n_{T}(\bx, \nu) \nonumber \\
Q_{\nu}(\bx) &=&  Q_{\rm CMB} (\bx) +g_{ \tau}(\bx) Q_{p}a_{\tau}+ n_{Q}(\bx, \nu) \nonumber \\
U_{\nu}(\bx) &=&  U_{\rm CMB} (\bx) + g_{ \tau}(\bx) U_{p} a_{\tau}+ n_{U}(\bx, \nu)  \label{eq:data_model}
\ea}
Here $Q_{p}$ and $U_{p}$ are the Stokes parameters describing the cosmological signal generated by remote quadrupole scattering. \TIB{They are constant in the patch surrounding the cluster}.  $n_{T}$, $n_{Q}$ and $n_{U}$ represent the instrumental noise in temperature and polarization data, with \TIB{$\sigma(n_Q)=\sigma(n_U)= \sqrt{2}\sigma(n_T) $}.
A derivation of the matched filter for the tSZ emission can be found in \cite{2016PhRvD..94d3522A}. The maximum-likelihood solution for the tSZ amplitude is given by
\ba
\frac{a^{\rm ML}_{\rm tSZ}}{ \sigma^{2}(a_{\rm tSZ})} &=&  \sum_{\nu, \nu'}  \int d\bl  g^{*}_{\rm tSZ}(\bl)  f_{\rm tSZ}(\nu)  [C^{-1}_{T} (\ell)]_{\nu, \nu'} T_{\nu'}(\bl)\nonumber \\
\frac{1}{\sigma^{2}(a^{\rm ML}_{\rm tSZ})} &=& \sum_{\nu, \nu'} f_{\rm tSZ}(\nu)  f_{\rm tSZ}(\nu')  \int d\bl  |g_{\rm tSZ}(\bl)|^{2}  [C^{-1}_{T} (\ell)]_{\nu, \nu'}  \nonumber \\
\ea
The noise covariance matrix is obtained by summing the background CMB power spectrum and the effective instrumental noise power spectrum
\ba
[C_{T} (\ell)]_{\nu, \nu'}&=& C_{TT}(\ell) + \delta_{\nu, \nu'} \frac{N_{\nu} (\ell)}{B^{2}_{\nu} (\ell)} \\
&=&   C_{TT}(\ell) + \delta_{\nu, \nu'} \tilde{N}_{\nu}
\ea
where $B_{\nu}(\ell)$   the frequency-dependent beam transfer function and $\tilde{N}_{\nu}$ is the effective noise power spectrum. 

The matched filter for the optical depth amplitude takes a similar form but with the additional complexity of having to consider the two Stokes parameters. A detailed derivation is presented in Appendix A. The maximum-likelihood solution for the optical depth is given by
\ba
\frac{a^{\rm ML}_{\tau}}{\sigma^{2}(a_{\tau})} &=&   \int d\bl  g^{*}_{\tau}(\bl) \begin{pmatrix}  Q_{p} \cr U_{p}  \cr \end{pmatrix}^{T} \nonumber  [C^{-1}_{P} (\ell)]  \begin{pmatrix}  Q(\bl) \cr U(\bl)  \cr \end{pmatrix} \nonumber \\
\frac{1}{\sigma^{2}(a_{\tau})} &=&   \int d\bl    \frac{ |g_{\tau}(\bl)|^{2} |p|^{2} }{C_{EE}(\ell)+ 2(\sum \tilde{N}^{-1}_{\nu}(\ell))^{-1}}   
\ea
\TIB{where $\bl$ is a two dimensional wavevector and $|p|=\sqrt{Q^{2}_{p}+U^{2}_{p}}$ is the norm of the cosmological signal}.  For forecasting  the signal-to-noise for an individual cluster we will use $ |p| \sim |p_{c}| \sim  \langle |p_{c} (z)| \rangle_{\rm S4} $ and neglect  $ p_{u}$, the  part of the remote signal generated by  remote quadrupole scattering but uncorrelated with our CMB measurement. For a single cluster, the variance produced by this term is sub-dominant compared to other sources of errors. It will be included in Section \ref{sec:Forecast} when we will consider the use of the full cluster catalog to constrain the relationship between optical depth and mass.
\subsection{CMB-S4 cluster catalog}\label{subsec:catalogl}

The uncertainties on the amplitude of the tSZ effect after matched filtering of the data can be used to predict the number of clusters detected by CMB S4. The  CMB S4 instrumental specifications assumed for this work are presented in Table \ref{tab:cmbtable}; they correspond to $\sim 10^{5}$ detectors and the angular resolution obtained with a three-meter mirror.

We follow \cite{2016PhRvD..94d3522A} and compute the cluster detection efficiency
\ba \label{eq:detecteff}
  \tilde{\chi}(M_{500},z) &=&\int d(\ln Y^{\rm true}_{500}) \int_{q\sigma_N}^\infty  dY_{500}^{\rm obs}  \\
  &&P_{\rm SZ}(\ln Y^{\rm true}_{500}|M_{500},z)  P_{\rm det}(Y_{500}^{\rm obs}|Y_{500}^{\rm true}) \nonumber 
\ea
\TIB{Here $P_{\rm SZ}$ accounts for the intrinsic scatter in the relationship between tSZ flux and mass (e.g \cite{2016PhRvD..94d3522A})}. $P_{\rm det}$  quantifies the uncertainty in the measurement of the tSZ amplitude for the CMB S4 specification. \TIB{We use a detection threshold of $q\sigma_{N}$,  $\sigma_{N}$ being the noise on the measurement of $Y_{500}$. In this work, we will only consider clusters detected with a tSZ signal-to-noise threshold $q > 6$. }

\begin{table}
  \centering{
  \renewcommand*{\arraystretch}{1.2}
  \begin{tabular}{|c|c|c|}
  \hline
  Frequency & Noise RMS & Beam FWHM \\
  (GHz) & ($\uKam$) & (arcmin) \\
  \hline
  28 & 9.8 & 14.0  \\
  41 & 8.9 & 10.0  \\
  90 & 1.0 & 5.0   \\
  150 & 0.9 & 2.8 \\
  230 & 3.1 & 2.0 \\
  \hline
  \end{tabular}}
  \caption{Instrumental noise and angular resolution of the 5 frequency channels of CMB S4. The final design of CMB S4 is still being discussed so these numbers should be taken with caution. In our analysis, we will only consider the 41, 90 and 150 GHz frequency channels and assume that the 28 and 230 GHz channel will be used as templates to clean the synchrotron, dust, and cosmic infrared background signal.}
  \label{tab:cmbtable}
\end{table}

 \begin{figure}
  \centering
  \includegraphics[width=1\columnwidth]{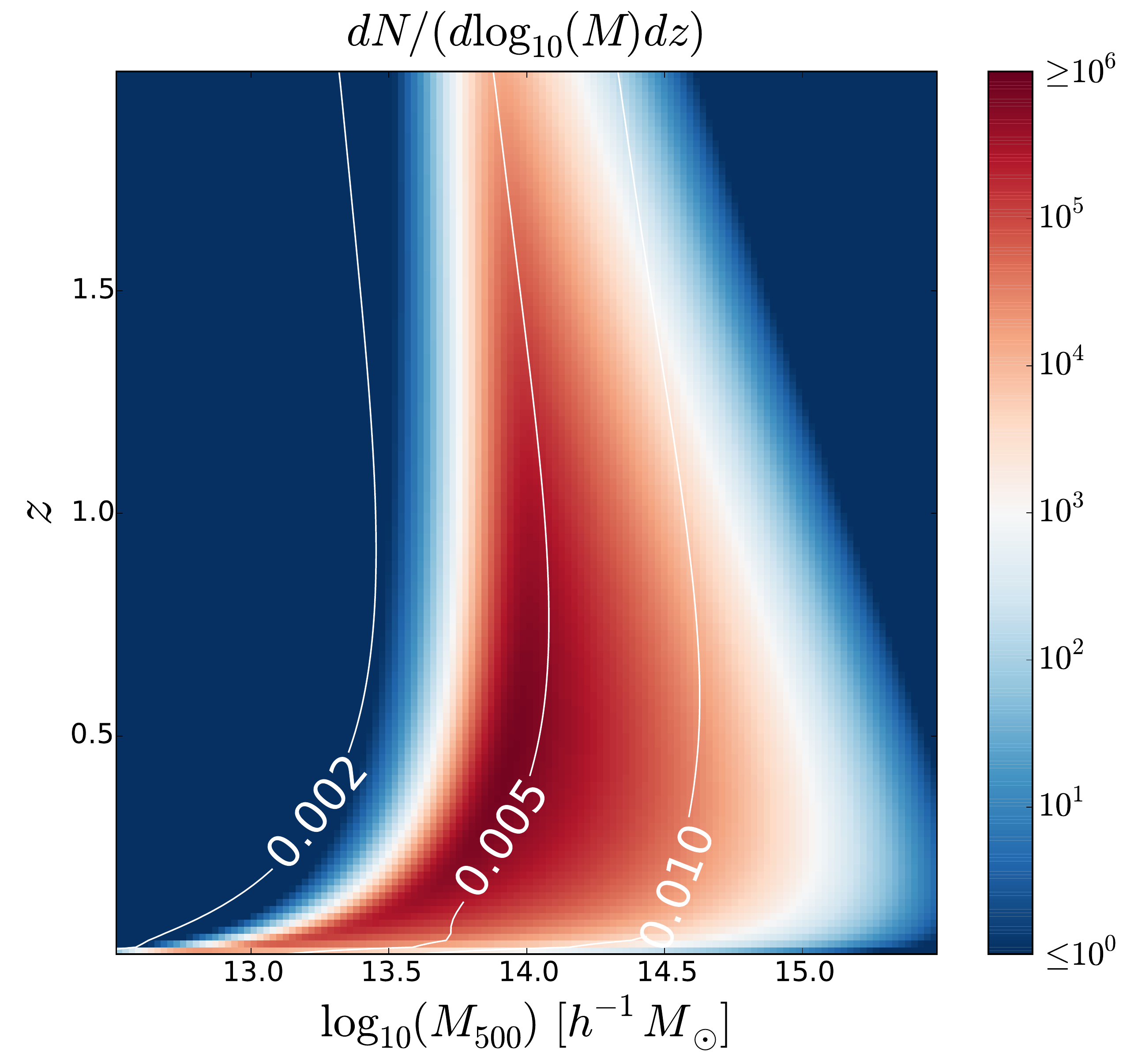}
  \caption{ \TIB{ Expected redshift and mass distributions for tSZ-selected clusters detected with an S4 experiment}. The white contours represent the signal-to-noise on the cluster polarization signal for individual clusters. The S/N for a typical cluster detected by CMB-S4 is small for the CMB-S4 specification.  }
  \label{fig:cluster_S_to_N}
\end{figure}

The S4 cluster catalog can then be obtained by multiplying the detection efficiency by the halo mass function. We display in Figure \ref{fig:cluster_S_to_N} the cluster distribution together with the signal-to-noise \TIB{($a_{\tau}/ \sigma(a_{\tau} $))}  for the individual clusters of the catalog.  The S/N  per individual cluster is extremely small, $0.5\%$ for a typical cluster in our catalog. Note that the E-mode background contributes to a strong degradation of the signal-to-noise.

\section{Relationship between optical depth and mass}\label{sec:Forecast}

While the detection of the polarized signal for an individual cluster is well beyond the reach of a CMB S4 experiment, the signal could be detected statistically using the full  S4 cluster catalog, and  could in principle be used to learn about the optical depth of the cluster population. Hydrodynamical simulations of cluster suggest a simple relationship between cluster optical depth and mass  \cite{2016JCAP...08..058B}. In this section, we first discuss the form of the Fisher matrix for the parameters describing this $\tau_{500}$-$M_{500}$ relationship. We then forecast constraints on these parameters using the CMB S4 instrumental specifications. For completeness, we also discuss the effect of increasing the angular resolution of CMB S4 on these constraints.
 
\subsection{Fisher Matrix}

We  use a power-law parametrization for the  $\tau_{500}$-$M_{500}$ relationship \TIB{
\ba
\tau_{500}=a_{\tau} (A,b) = A \tau_{*} \left( \frac{d_A(z_{*})}{d_{A}(z)}\right)^{2} \left( \frac{M_{500}}{M_{*}} \right)^{b}
\ea }
 with the pivot mass $M_{*}=1.2\times 10^{14} h^{-1} M_{\odot}$\TIB{, pivot redshift $z_{*}=0.5$,} and fiducial values $(A,b)=(1,1)$. The Fisher matrix for the full cluster catalog can be written
\ba
 F_{\alpha \beta} &=& \sum_{ij} \frac{ \partial a_{\tau,i}  }{ \partial \alpha} p_{c,i} (C^{-1})_{ij}p_{c,j}  \frac{ \partial a_{\tau,j}  }{ \partial \beta}  \\
 &+& \frac{1}{2} {\rm Tr}( C^{-1}C_{,\alpha}  C^{-1}C_{,\beta}) \nonumber
 \ea
 where the sum over $i$ and $j$ is taken over all clusters from the catalog.
The covariance matrix has contributions from two terms: an error  for each individual cluster measurement and an error term coming from the part of the remote quadrupole signal uncorrelated with our local CMB measurement 
\ba
C_{ij}= \sigma^{2}_{i} \delta_{ij} +  a_{\tau,i}  a_{\tau,j}\xi^{U}_{ij}.
\ea
\TIB{
For a cluster of mass $M$ 
at redshift $z$
\ba
\sigma^{-2}_i =  \frac{1}{8\pi \theta^{2}_{500}}  \int     \frac{ dk k |u(k)|^{2} }{(C_{EE}(k/\theta_{500})+2\tilde{N}(k/\theta_{500}))}  
\ea
This derivation of this equation is equivalent to the one presented in (\ref{eq:filtervariance}) of  Appendix A, but with the polarisation amplitude $|p^{2}|$  factorized out. }
The interpretation of the Fisher matrix is simple. The first term quantifies the constraining power in the change in the correlated polarized emission due to a change of the optical depth of the cluster population. The  polarized emission can be directly compared with the expected emission inferred from measurement of our local last scattering surface. The trace  accounts for the constraining power on the optical depth in the uncorrelated component of the polarized emission. This second term carries very little information so we can safely neglect it. An analytical computation of the Fisher matrix is possible: using the fact that the coherence length of the cosmological signal is much larger than the angular extent of the galaxy clusters, we get
\TIB{
\ba \label{eq:fishText}
F_{\alpha \beta}&=& \frac{1}{4\pi}  \int dz f_{\alpha \beta}(z) \sum_{\ell m} p_{c, \ell m}(z) p^{*}_{c,\ell m} (z) \nonumber\\
&&-  \frac{1}{4\pi}  \int dz dz' f_{\alpha}(z) f_{\beta}(z') \nonumber\\
&&\qquad\sum_{\ell m}   p_{c, \ell m}(z)[C_{\ell}(z,z')]^{-1} p^{*}_{c, \ell m}(z')   
 \ea}
The derivation of this result and the form of the weight factors $f_{\alpha \beta}(z)$ and $f_{\alpha}(z)$ are provided in the Appendix. The first term of this expression corresponds to a simple inverse-variance weighting combination of all individual cluster measurements, and the second term accounts for the increase in variance due to the uncorrelated part of the remote quadrupole signal.  We note that the Fisher matrix does not include uncertainties on the mass measurement of the cluster. One possibility to infer cluster masses is to use the relationship between  tSZ flux and cluster mass (e.g. \cite{2016A&A...594A..24P}), which could be accurately calibrated using the high signal-to-noise measurement of CMB lensing \cite{2017PhRvD..95d3517L,2015A&A...578A..21M}. The uncertainties on cluster masses will always be subdominant compared to  measurement errors on the polarization signal. \TIB{This fisher matrix does not take into account correlated noise between different cluster optical depth measurements. This assumption will break down for nearby cluster due to the correlation lenght of the background E modes.}

\begin{figure}
  \centering
  \includegraphics[width=1\columnwidth]{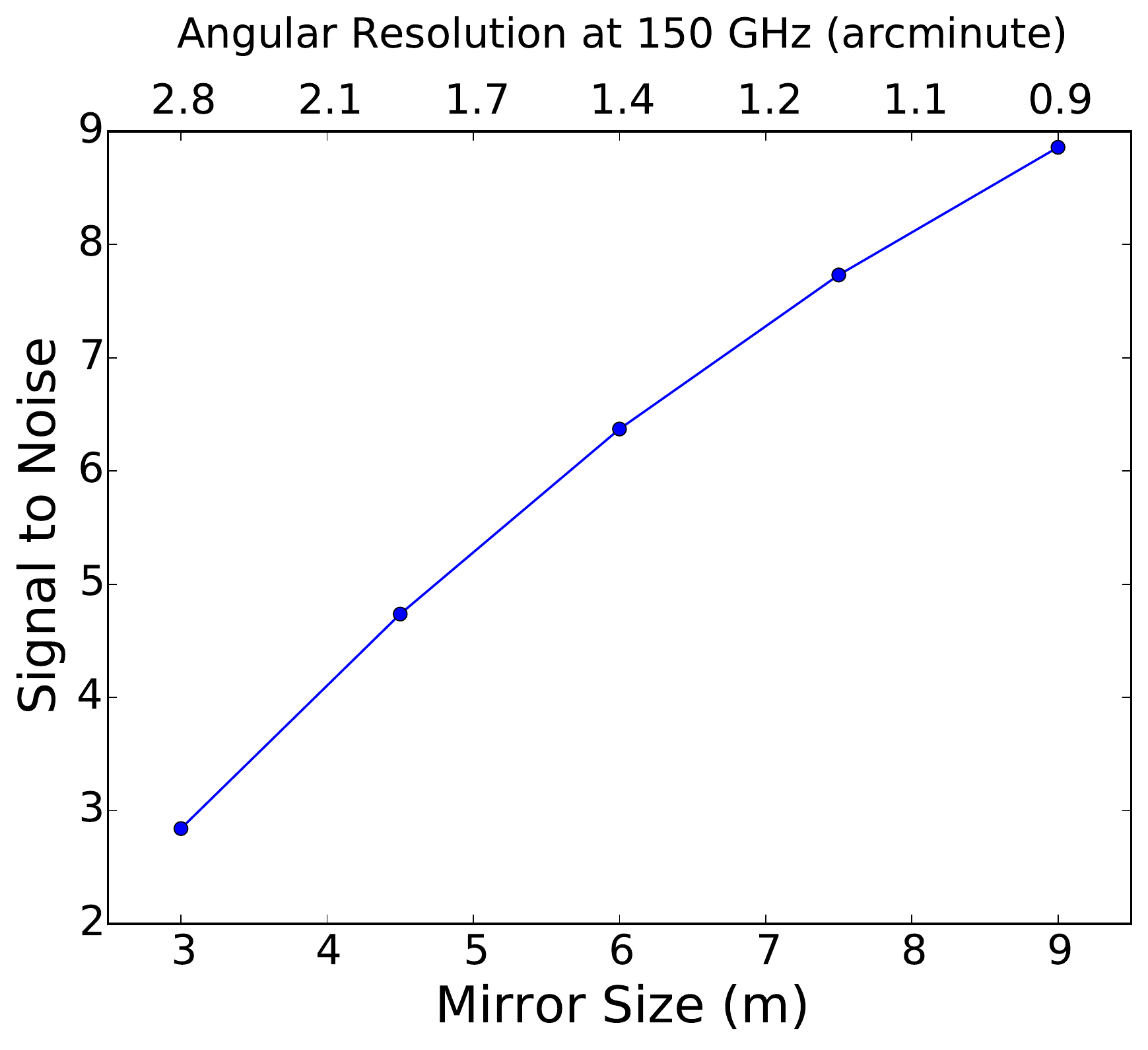}
  \caption{Aggregate S/N on the measurement of the cluster polarized emission as a function of the mirror size for a future CMB S4 experiment. The S/N is around 3 for the fiducial S4 specifications and scales roughly linearly with mirror size. \TIB{Improvement on the S/N is due to the increased number of clusters and the reduced effective noise of the matched filter for each cluster. The second effect dominates at low and intermediate resolution while the first effect becomes important for telescope mirror $>7 {\rm m}$.}}
  \label{fig:S_to_N_full}
\end{figure}

 \begin{figure}
  \centering
  \includegraphics[width=1\columnwidth]{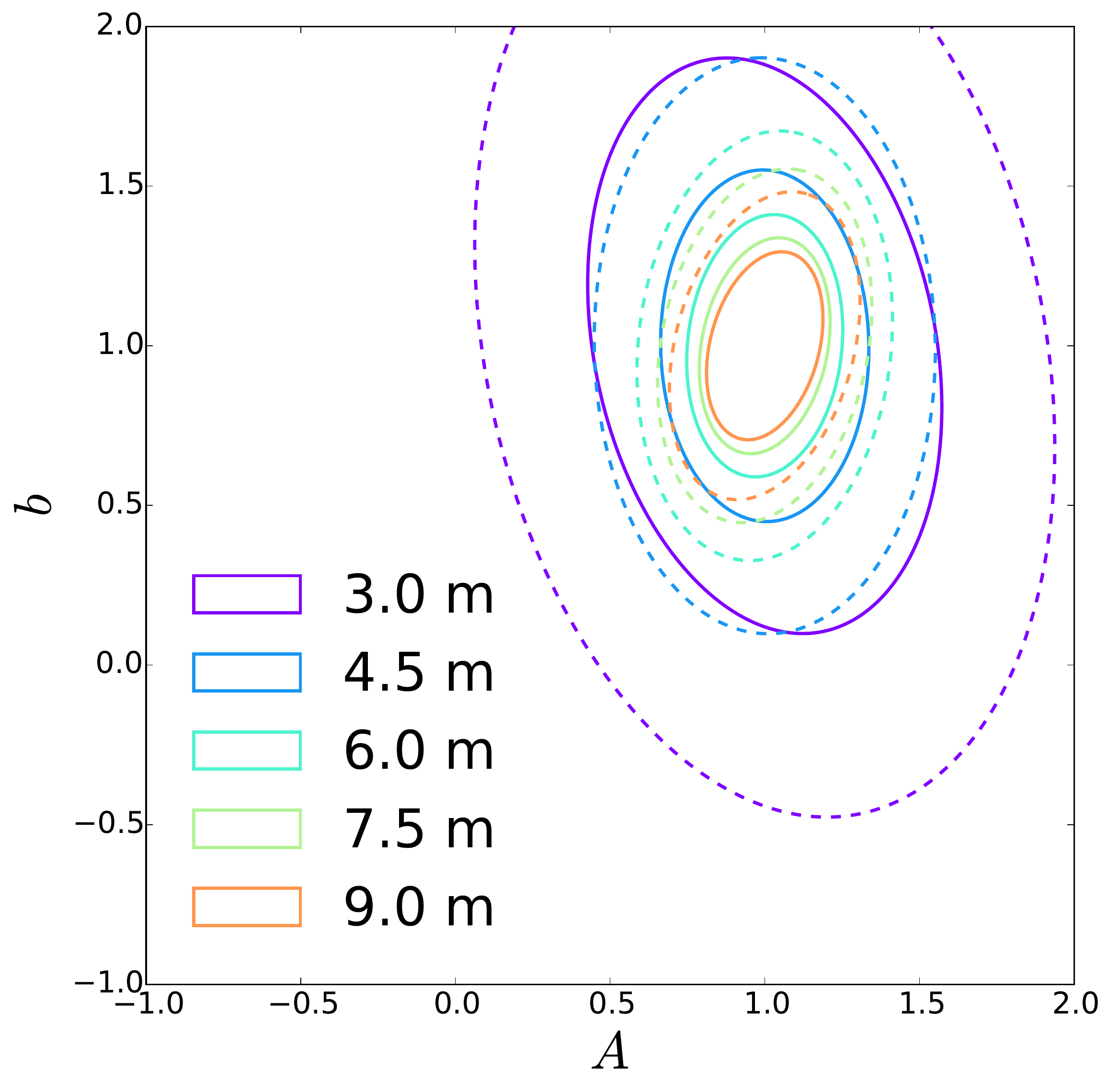}
  \caption{ 68\% and 95\% confidence levels on the power-law parameters of the relationship between optical depth and mass. The results suggest that the baseline specifications of CMB S4 with 3-meter mirrors might not be enough to calibrate the relationship using measurement of the cluster polarization signal. Increasing the mirror size helps by increasing the number of detected clusters and by reducing the uncertainties on the matched filter for individual clusters.}
  \label{fig:tau_M}
\end{figure}

\subsection{Result for different CMB S4 experimental specifications}

The Fisher matrix allows us to forecast the expected error bars on the power law parameter $A$ and $b$ of the $\tau_{500}$-$M_{500}$ relationship, using the constraining power in the correlated remote quadrupole signal. 
We choose as a baseline the angular resolution displayed in  Table \ref{tab:cmbtable}, which corresponds to the angular resolution achievable with three-meter mirrors, but we also investigate the effect of increasing the mirror size. The results of this analysis are shown in  Figure \ref{fig:S_to_N_full} and Figure \ref{fig:tau_M}.  Figure \ref{fig:S_to_N_full} shows the total signal to noise on the correlated polarized emission, and Figure \ref{fig:tau_M} shows the 68\% and 95\% confidence levels on the $\tau_{500}$-$M_{500}$ relationship parameters. For the fiducial CMB S4 specifications, we expect a $S/N \sim 3$,  which would not result in a useful characterization of the $\tau_{500}$-$M_{500}$ relationship. The S/N improves linearly with the telescope diameter. For example, if S4 is composed of telescopes observing at arcminute resolution in the 150 GHz band (corresponding to a 9-meter mirror)  it will reach a signal-to-noise of around nine on the cluster polarized emission. 

\section{Discussion} \label{sec:conclusion} 
In this paper we have investigated the possibility for the next-generation CMB S4 experiment to detect the polarized emission generated by remote quadrupole scattering on the hot electron gas inside clusters. We find that this detection would be difficult, with a signal to noise of only 0.5$\%$ for typical cluster detected by S4 and an overall expected detection of only 3$\sigma$  for the CMB S4 fiducial specifications. We find that the signal-to-noise will increase linearly with respect to the angular resolution of the telescope, reaching a 9$\sigma$ detection for a 9 meter mirror.  

We also discuss the possibility of using the signal to calibrate the relationship between optical depth and mass of the cluster, using the strong correlation between the polarized emission at low redshift and our own last scattering surface observations. 

We should note that Fisher forecasts tends to be on the optimistic side. While using the matched filter formalism we implicitly had to assume perfect knowledge of the cluster spatial profiles. This allowed us to optimally separate the cluster signal from the background CMB polarization. However, in practice, uncertainties on these profiles will also increase the uncertainties of the detection. 

In this paper, we consider quadrupole scattering in galaxy clusters, treating each cluster as a single measurement of the cosmological signal $p$, applying a matched filter and integrating over the cluster area. We can instead imagine measuring the modulated signal ${\cal P}=\tau p$ for a spatially-varying optical depth $\tau$. The cosmological signal $p$ caused by quadrupole scattering contains only E modes (for scattering at a fixed redshift), but the observed signal ${\cal P}$ contains B modes due to modulation by the spatially-varying optical depth $\tau$ \cite{2017arXiv171001708M}. The B component of ${\cal P}$ may be of interest, because the B-mode background is much smaller than the E-mode background, so the signal to noise on the B-mode polarized emission generated by remote quadrupole scattering could eventually exceed that of the E-mode field. Investigation of the properties of this signal is left to future work. 

To conclude, CMB S4 could achieve a significant detection of the cluster signal if it is made of high-resolution telescopes. The signal could then be used to get a first calibration of the $\tau_{500}$-$M_{500}$ relationship. 
As shown in Appendix C, the exploitation of the full potential of the cluster polarization signal will require improving the noise level on the CMB sky even beyond CMB S4 specifications.  A combination of the cluster polarization measurement with the kSZ measurement could then be used to test the $\Lambda$CDM model while keeping under control baryonic physics affecting the cluster optical depth. 

\section*{Acknowledgments}
We thank David Alonso, Sigurd Naess and Alexander van Engelen  for useful discussions. TL and BW are supported  by the
Labex ILP (reference ANR-10-LABX-63) part of the Idex
SUPER, and received financial state aid managed by
the Agence Nationale de la Recherche, as part of the
programme Investissements d'avenir under the reference
ANR-11-IDEX-0004-02. TL and J.S. has been supported in part by ERC Project No. 267117 (DARK) hosted by the Pierre and Marie Curie University-Paris VI, Sorbonne Universities. EFB is supported by NSF Award 1410133. 

 \bibliography{draft}

\appendix
\section{Matched filter for the optical depth parameter}

In this appendix, we present the derivation of the matched filter for the optical depth parameter $a_{\tau}$ for a single cluster.
We start with the log-likelihood for $a_{\tau}$ at a single frequency
\ba
\chi^{2} &=&  \int d\bl  \bm{V}^{T}(\bl) [C^{-1}_{P} (\bl)] \bm{V}(\bl) \\
 \bm{V}(\bl) &=& \begin{pmatrix}  Q(\bl) \cr U(\bl)  \cr \end{pmatrix} -  g_{ \tau}(\bl) a_{\tau} \begin{pmatrix}  Q_{p} \cr U_{p}  \cr \end{pmatrix} \nonumber
\ea
Here $Q$ and $U$ are the observed Stokes parameters, and $Q_{p} $ and $ U_{p} $ represent the cosmological signal generated by remote quadrupole scattering.
The noise covariance matrix has contributions from the background Q,U polarization field and from instrumental noise
\ba
[C_{P} (\bl)]= \begin{pmatrix} 
C_{QQ}(\bl) & 
C_{QU}(\bl) \cr
C_{QU}(\bl)& 
C_{UU}(\bl)  \cr
\end{pmatrix} 
+ 2  \tilde{N}(\ell) \textrm{I}_{2\times 2},
\ea 
where $I_{2\times 2}$ is the two-dimensional identity matrix and $N_{UU}(\ell)=N_{QQ}(\ell)=2N_{TT}(\ell)$.

It is convenient to transform the Stokes parameters to E and B modes. In a small patch around the cluster (e.g. \cite{2013MNRAS.435.2040L})
\ba
E({\bl}) \pm i B({\bl})&=& e^{\mp 2i\phi_{\bl}}(Q(\bl) \pm i U(\bl))
\ea
Here $\phi_{\ell}$ is the angle between the Fourier wave vector $\bl$ and $\ell_{x}$ the $x$ axis of the Fourier plane. Assuming that the primordial and lensed B modes can be neglected with respect to  \TIB{ the sum of the E-mode background and instrumental noise}, we have
\ba
C_{QQ}(\bl) &=& C_{EE}(\ell) \cos^{2} 2 \phi_{\bl} \nonumber \\
C_{QU}(\bl) &=& C_{EE}(\ell) \cos 2 \phi_{\bl}  \sin 2 \phi_{\bl} \nonumber \\
C_{UU}(\bl) &=& C_{EE}(\ell) \sin^{2} 2 \phi_{\bl} 
\ea
The maximum-likelihood solution is found by setting the derivative of the log-likelihood with respect to the amplitude parameter to zero, yielding
\ba
\frac{a^{\rm ML}_{\tau}}{\sigma^{2}(a_{\tau})} &=&   \int d\bl  g^{*}_{\tau}(\bl) \begin{pmatrix}  Q_{p} \cr U_{p}  \cr \end{pmatrix}^{T} \nonumber  [C^{-1}_{P} (\ell)]  \begin{pmatrix}  Q(\bl) \cr U(\bl)  \cr \end{pmatrix}  \\
\frac{1}{\sigma^{2}(a^{\rm ML}_{\tau})} &=&   \int d\bl  |g_{\tau}(\bl)|^{2} \begin{pmatrix}  Q_{p} \cr U_{p}  \cr \end{pmatrix}^{T}  [C^{-1}_{P} (\ell)] \begin{pmatrix}  Q_{p} \cr U_{p}  \cr \end{pmatrix}  \nonumber \\
\ea
After a bit of algebra we can simplify the expression of $\sigma^{2}(a^{\rm ML}_{\tau})$
\ba
\begin{pmatrix}  Q_{p} \cr U_{p}  \cr \end{pmatrix}^{T}  [C^{-1}_{P} (\ell)] \begin{pmatrix}  Q_{p} \cr U_{p}  \cr \end{pmatrix} = \frac{ |p^{2}| 2\tilde{N}(\ell) + C_{EE}(\ell)B^{2}_p}{2\tilde{N}(\ell)(C_{EE}(\ell)+2\tilde{N}(\ell))} \nonumber \\
\ea
with $|p^{2}|=Q^{2}_{p}+U^{2}_{p}$, and $B^{2}_{p}$ the B modes part of the cosmological signal generated by remote quadrupole scattering. 
Note that this term has a simple interpretation. If the polarization generated by remote quadrupole scattering was purely B mode (with $B^{2}_{p}= |p^{2}|$), we would be able to distinguish it from the E mode background, and the variance of the filter would be purely given by the variance of the instrumental noise. \TIBO{The most conservative  scenario is obtained by setting $ B_{p}=0$, and the variance gets contributions both from the E mode background and the instrumental noise}
\ba
\frac{1}{\sigma^{2}(a_{\tau})} &=&   \int d\bl    \frac{ |g_{\tau}(\bl)|^{2} |p^{2}| }{(C_{EE}(\ell)+2\tilde{N}(\ell))}   .
\ea
\TIB{Our approach is thus local, and the signal is measured cluster by cluster. Recently, nonlocal approaches have been proposed, which use the fact that  B modes can be generated from the modulation of the remote scattering signal by the free electron density across the entire Universe (see for example \cite{2017arXiv170508907D}, \cite{2017arXiv171001708M} and \cite{2017arXiv170708129D}  ).}
For multifrequencies observations we replace the effective noise covariance matrix by the minimum variance combination:
\ba
\frac{1}{\tilde{N}(\ell)}= \sum_{\nu} \frac{1}{\tilde{N}_{\nu}(\ell)}
\ea
We can simplify this expression even further, using the azimuthal symmetry of the profile. Its Fourier transform 
\ba
g_{\tau}(\bl)&=& \frac{1}{4\pi \theta^{2}_{500}} \int \frac{d\bx}{2\pi} e^{i \bl .\bx} u_{\tau}(\bx) \\
u_{\tau}(\bx) &=&  \frac{\int_{-\infty}^{\infty} dx_{z} \Gamma_{ \tau} ( \sqrt{x_{z}^{2} + x^{2}}) }{  \int^{1}_{0}dx_{r} x^{2}_{r}\Gamma_{ \tau}(x_{r}) }
\ea
is given simply by \TIB{
\ba
g_{\tau}(\bl)&=&\frac{1}{4\pi}u_{\tau}(k=\ell \theta_{500}) \\
u_{\tau}(k=\ell \theta_{500})&=&  \int dx x J_{0}(\ell \theta_{500} x)u_{\tau}(x) 
\ea}
 The covariance becomes
\ba
\frac{1}{\sigma^{2}(a_{\tau})} &=&  \frac{1}{8\pi \theta^{2}_{500}}  \int     \frac{ dk k |u(k)|^{2} |p^{2}| }{(C_{EE}(k/\theta_{500})+2\tilde{N}(k/\theta_{500}))}   \nonumber  \label{eq:filtervariance}\\
\ea

\section{Analytical form of the Fisher Matrix}

The dominant term of the Fisher matrix for the cluster polarized emission can be written as
\ba
F_{\alpha \beta} &=& \sum_{ij} \frac{ \partial a_{\tau,i}  }{ \partial \alpha} p_{c,i}  (C^{-1})_{ij} p_{c,j}  \frac{ \partial a_{\tau,j}  }{ \partial \beta} 
\ea
with covariance matrix
\ba
C_{ij}= \sigma^{2}_{i} \delta_{ij} +  a_{\tau,i}  a_{\tau,j}\xi^{U}_{ij}
\ea
The sum over $i$ and $j$ is taken over the 200 000 clusters of the CMB S4 catalog. The coherence length of the quadrupole can be used to reduce the number of degrees of freedom in the problem.
We can define a set of voxels, volume elements in which the cosmological signal is constant, and use a projection operator between voxel and cluster 
\ba
p_{c,i} &=& \sum_{v} J_{iv} p_{c,v}  \\
\xi^{U}_{ij} &=&\sum_{vw} J_{iv} \xi^{U}_{vw} J_{wj}
\ea 
where $J_{iv}$ is unity if the cluster $i$ belong to the voxel $v$ and zero otherwise. The covariance matrix then becomes 
\ba
C_{ij} &=& \sigma_{i}^{2} \delta_{ij} +  a_{\tau,i}  a_{\tau,j}\sum_{vw} J_{iv} \xi^{U}_{vw} J_{wj}  
\ea
and can be inverted using the Woodbury formula, 
\TIBO{
\ba
(C^{-1})_{ij} &=&  \sigma_{i}^{-2} \delta_{ij} -  \sum_{vw} \frac{ a_{\tau,i}} {\sigma_{i}^{2}} J_{iv}  (M^{-1})_{vw}  J_{wj}   \frac{a_{\tau,j}} {\sigma_{j}^{2}} \nonumber \\  
M_{vw} &=&  \left(  ( \xi^{-1})_{vw}^{U} +  \delta_{vw} \sum_{k \in v} \frac{a^{2}_{\tau,k}}{\sigma^{2}_{k}}  \right)
\ea
}
This allows us to get an analytic estimate of the Fisher matrix
\ba \label{eq:fishApp}
F_{\alpha, \beta}   &=&  \sum_{v} |p_{c,v}|^{2} f_{\alpha \beta, v}   \\ &-& \sum_{vw}  f_{\alpha,v} p_{c,v}  (M^{-1})_{vw}  p^{*}_{c,w}  f_{\beta,w} \nonumber
 \ea
 The weight factors $f_{\alpha \beta, v} $ and $f_{\alpha}$ can be written as integrals over the cluster distribution
 \begin{widetext}
 \ba
 f_{\alpha \beta, v} &=&  4\pi \int_{z_{v}}^{z_{v}+\delta z} f^{*}_{\rm sky}(z)  dz  \frac{r^{2}(z)}{H(z)} \int dM \frac{ n(M,z)\tilde{\chi}(M,z)}{{\sigma^{2}(M,z)}} \frac{\partial a(M,z) }{\partial \alpha}  \frac{\partial a(M,z) }{\partial \beta} \\
f_{\alpha, v} &=&  4\pi  \int_{z_{v}}^{z_{v}+\delta z} f^{*}_{\rm sky}(z) dz  \frac{r^{2}(z)}{H(z)} \int dM \frac{ n(M,z)\tilde{\chi}(M,z)}{{\sigma^{2}(M,z)}} a(M,z)  \frac{\partial a(M,z) }{\partial \alpha} \\
\TIB{\sigma^{-2}(M,z)} &=&  \frac{1}{8\pi \theta^{2}_{500}}  \int     \frac{ dk k |u(k)|^{2} }{(C_{EE}(k/\theta_{500})+2\tilde{N}(k/\theta_{500}))}
\ea
\end{widetext}

 \begin{figure}
  \centering
  \includegraphics[width=1\columnwidth]{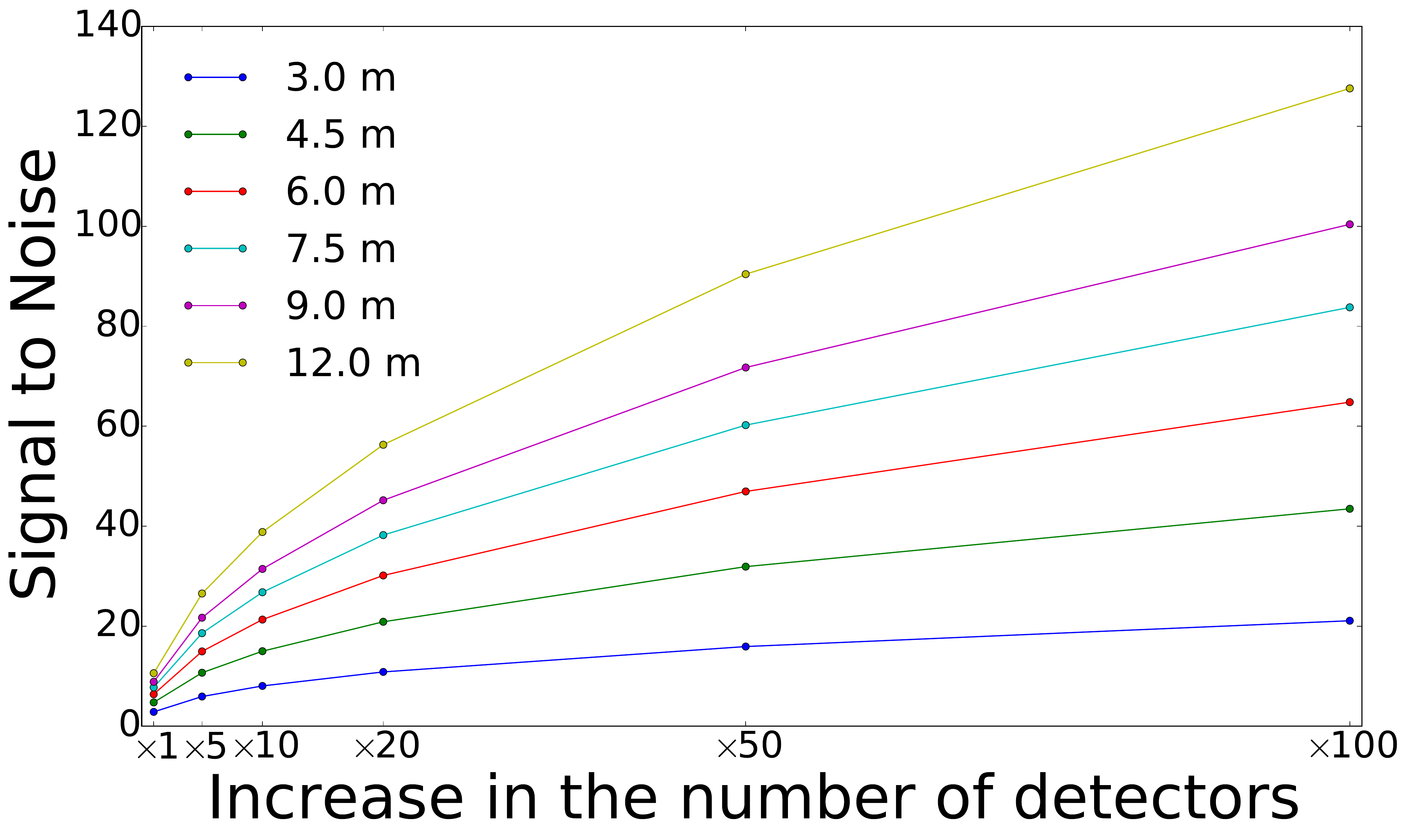}
  \caption{S/N on the cluster polarization signal as a function of number of detectors and mirror size. The baseline number of detector corresponds to CMB S4: 100 000 detectors. These forecasts do not include the B modes signal generated by the modulation of the E modes signal by the electron density field. They should be seen as pessimistic. }
  \label{fig:S_to_N_all}
\end{figure}
where $n(M,z)$ is the halo mass function, $\tilde{\chi}(M,z)$ is the detection efficiency (see Equation \ref{eq:detecteff}), and $\sigma^{2}(M,z)$ is the error on the polarized emission for a single cluster which depends only on the angular extent of the cluster $\theta_{500}(M,z)$. Note that  $f_{\alpha \beta, v}$ and $f_{\alpha, v}$  depend only on the redshift of the voxels and not on their angular position on the sky. However, the signal is slightly anisotropic and most of the constraining power is located in the S4 survey area (see Section \ref{sec:Pol}).  We take this into account by using an effective sky fraction $f^{*}_{{\rm sky}}(z)= f_{{\rm sky}} \langle |p^{\rm S4}_{c}(z) | \rangle/ \langle |p^{\rm full}_{c} (z)| \rangle $. Equation \ref{eq:fishApp} is then equivalent to Equation \ref{eq:fishText} after a spherical harmonic transform. \\

\section{Beyond CMB S4}
The paper focusses on the possible detection and exploitation of the polarized signal emitted by clusters with the upcoming CMB S4 experiment. CMB S4 consists of $\sim 10^{5}$ detectors observing at microwave frequencies and is expected  to reach noise level of 1 $\mu$K arcmin over half the sky. In Figure \ref{fig:S_to_N_all} we show the  signal-to-noise improvement for more futuristic experiments. We should note that using a matched filter to extract the cluster polarization signal requires the assumption that clusters are isolated objects; for these futuristic experiments, this assumption will break down and  cluster blending will become important. Another important signal for the next generation experiment is the B modes signal arising from the modulation of the remote quadrupole scattering E modes field by the electron density field. This signal might be detected with a higher signal-to-noise as it is not degenerate with the strong primordial E modes background. The forecasts presented in this section should then be seen as  pessimistic.

\end{document}